\documentclass[twocolumn,prl,aps,superscriptaddress,footnote,amsfonts,showpacs]{revtex4-1}
\usepackage{amsmath}
\usepackage{amssymb}
\usepackage{bm}
\usepackage{epsfig}
\usepackage{color}
\newcommand{\bef}{\begin{figure}[htb]\centering}
\newcommand{\eef}{\end{figure}}
\newcommand{\ben}{\begin{eqnarray}}
\newcommand{\een}{\end{eqnarray}}
\newcommand{\nnu}{\nonumber\\}

\begin{document}
\title{Indication on the process-dependence of the Sivers effect}

\author{Leonard Gamberg}
\affiliation{Division of Science, 
                   Penn State University-Berks, 
                   Reading, PA 19610, USA}

\author{Zhong-Bo Kang}
\affiliation{Theoretical Division, 
                   Los Alamos National Laboratory, 
                   Los Alamos, NM 87545, USA}

\author{Alexei Prokudin}
\affiliation{Theory Center, Jefferson Lab, 
                   12000 Jefferson Avenue, 
                   Newport News, VA 23606, USA}

\begin{abstract}
We analyze the spin asymmetry for single inclusive jet production in proton-proton collisions
collected by AnDY experiment at the Relativistic Heavy Ion Collider 
and the Sivers asymmetry data from semi-inclusive deep inelastic 
scattering experiments. In particular, we consider the role color gauge invariance plays in determining 
the process-dependence of the Sivers effect. We find that after carefully taking into account
the initial-state and final-state interactions between the active parton and the remnant of the polarized hadron,
the calculated jet spin asymmetry based on the Sivers functions extracted from HERMES and 
COMPASS experiments is consistent with the AnDY experimental data. 
This provides  a first 
indication for the process-dependence of the Sivers effect in these processes. We also make predictions for both direct photon and Drell-Yan spin asymmetry, 
to further test the process-dependence of the Sivers effect in future experiments.
\end{abstract}


\maketitle

The investigation of nucleon's sub-structure has entered a new era. 
In past decades an understanding of nucleons in terms of quarks and gluons (partons), the degrees of freedom of Quantum Chromodynamics (QCD), has been
successfully established.
Progress was achieved in constructing a ``one-dimensional'' light-cone
picture of the nucleon based on the longitudinal motion of 
partons in fast moving nucleons. In recent years 
theoretical breakthroughs extended this description in the 
transverse as well as light-cone momentum space (three dimensions). 
Transverse-spin dependent observables, such as single transverse spin 
asymmetries (SSAs) provide firm evidence  
for a  three-dimensional tomography of the nucleon due to a non-trivial correlation between the transverse spin and the parton's transverse momentum, and present unique opportunities to study QCD dynamics, particularly QCD factorization and universality of the parton distributions~\cite{EIC}.

Large SSAs have been measured in fixed-target and collider mode in single inclusive particle production in nucleon-nucleon scattering experiments~\cite{Adams:1991cs} and semi-inclusive deep inelastic lepton-nucleon scattering (SIDIS)  experiments~\cite{hermes, compass, Qian:2011py}. Two different yet related QCD factorization formalisms have been proposed to describe the asymmetries. One relies on the so-called transverse momentum dependent (TMD) factorization~\cite{Ji:2004wu,TMD}, which is valid for the processes with two characteristic scales; for example the photon's virtuality $Q$ and $P_{h\perp}$ of the produced hadron in SIDIS, where $ \Lambda_{\rm QCD}^2 \lesssim P_{h\perp}^2 \ll Q^2 $. 
In this formalism transverse spin effects are associated with  
TMD parton distribution functions and fragmentation 
functions (PDFs and FFs). Then there is the collinear factorization formalism at next-to-leading power (twist-3) in the hard scale~\cite{Efremov:1981sh,Kang:2010zzb}. This approach is valid for processes with only one characteristic hard scale, for instance, the transverse momentum $P_{h\perp}^2 \gg \Lambda_{\rm QCD}^2$ of the produced hadron in  proton-proton ($pp$) collisions. It describes the spin asymmetry in terms of twist-3 three-parton correlation functions. One of the well-known examples is the so-called Efremov-Teryaev-Qiu-Sterman (ETQS) function $T_{q, F}(x, x)$~\cite{Efremov:1981sh}.

Of central importance in the study of SSAs is the
Sivers~\cite{Sivers:1989cc}  effect which has 
attracted great attention in recent years. 
In part this is due to the unique prediction from TMD factorization theorems
that the Sivers effect is process-dependent: that is
its existence relies on the initial-state and final-state interactions (ISIs and FSIs) between the struck parton and the remnant of the polarized hadron. These interactions depend on the color flow of the specific scattering process considered, thus giving rise to process-dependent Wilson lines in the gauge-invariant definition of the relevant TMD PDFs - in this case so-called Sivers functions $f_{1T}^{\perp}(x, k_\perp^2)$. The often discussed case is the difference between the FSIs in SIDIS and the ISIs in Drell-Yan (DY) production in $pp$ collisions which leads to an opposite sign in the Sivers function probed in these two processes, indicating that the Sivers function is not universal~\cite{Collins:2002kn}.   

On the other hand in the twist-3 collinear factorization approach, the process-dependence of the ISIs and FSIs is absorbed into the short-distance, 
perturbative cut scattering amplitudes, where the 
relevant twist-3 three-parton correlation functions are universal.
 As a result, TMD and collinear twist-3 factorization formalisms are closely related to each other~\cite{Ji:2006ub}. The relevant functions - the Sivers function and the ETQS twist-3 function are connected 
through the following relation~\cite{Boer:2003cm,Kang:2011hk},
\ben
T_{q, F}(x, x)=-\int d^2k_\perp \frac{|k_\perp|^2}{M}f_{1T}^{\perp q}(x, k_\perp^2)|_{\rm SIDIS}.
\label{TF}
\een
where the subscript  emphasizes that the Sivers function is probed in the SIDIS process. In other words, starting from the Sivers functions extracted from SIDIS, one can derive a functional form for ETQS function $T_{q, F}(x, x)$. In 
combination with the calculable short-distance cut scattering amplitudes, 
one should be able to predict the SSAs of inclusive particle production in $pp$  collisions.  However, a recent study~\cite{Kang:2011hk,Kang:2012xf,Gamberg:2010tj} 
for inclusive {\it hadron} production in $pp$ collisions shows that such calculated SSAs are {\it opposite} to those measured in the experiments. This is known as the ``sign mismatch'' problem. 
Whether this finding reflects the inconsistency of our theoretical formalism is a very important question and needs to be explored both theoretically and experimentally. However, since the SSAs of inclusive hadron production can also receive contributions from the fragmentation process~\cite{Kang:2010zzb}, a thorough analysis demands including such contributions. 

A new opportunity presents itself however, with a recent inclusive jet measurements performed at the AnDY experiment at RHIC~\cite{AnDY}.  Since the jet spin asymmetry does not involve fragmentation contributions,  this paves the way  to
precisely test the process-dependence of the Sivers effect in 
different processes \cite{endnote} as well as explore the consistency of the TMD and collinear twist-3 factorization formalisms~\cite{Kang:2011hk,Metz:2012ui,D'Alesio:2011mc}. This is the main purpose of our paper.  
We analyze the spin asymmetry for single inclusive jet production in $pp$ collisions collected by the AnDY experiment and the Sivers asymmetry data from SIDIS experiments. We assess whether they are compatible with each other; in other words, whether the jet asymmetry is consistent with our expectation on the process-dependence of the Sivers effect. 

We start with the basic formalism for the SIDIS SSA. For hadron production in SIDIS at low transverse momentum $P_{h\perp}$, 
$e (\ell) + A^\uparrow(P, s_\perp)\to e (\ell') + h(P_h) + X$, within the TMD factorization formalism, 
the differential cross section for the Sivers effect reads~\cite{Bacchetta:2006tn},
\ben
\frac{d\sigma}{d\, \mathcal{P.S.}}
&=& \sigma_0 \left[F_{UU,T} +   \sin(\phi_h-\phi_s) F_{UT,T}^{\sin(\phi_h -\phi_s)} \right],
 \label{eq:aut}
\een
where phase space $d\, \mathcal{P.S.}=dx_B dy d\phi_s dz_h d\phi_h P_{h\perp} d P_{h\perp}$ with the standard SIDIS kinematic variable $x_B$, $y$, and $z_h$. The normalization factor $\sigma_0=\sigma_0(x_B, y, Q^2)$ and the structure functions $F_{UU,T}$ and $F_{UT,T}^{\sin(\phi_h -\phi_s)}$ are defined in Ref.~\cite{Kang:2012xf}. The Sivers asymmetry measured in the experiments is defined by
$A_{UT}^{\sin(\phi_h -\phi_s)}({\scriptstyle{x_B, z_h, P_{h\perp}}})=
\sigma_0({\scriptstyle{x_B, y, Q^2}}) F_{UT,T}^{\sin(\phi_h -\phi_s)}/
\sigma_0({\scriptstyle{x_B, y, Q^2}}) F_{UU ,T}$.

On the other hand, the single inclusive jet production in transversely-polarized $pp$ collisions, 
$A(P_A, s_\perp)+B(P_B)\to {\rm jet}(P_J)+X$, only receives the Sivers type of contributions. Within the collinear factorization formalism, the spin-dependent differential cross section $d\Delta\sigma(s_\perp) = \left[d\sigma(s_\perp)-d\sigma(-s_\perp)\right]/2$ can be written as,
\ben
E_J\frac{d\Delta\sigma(s_\perp)}{d^3P_J} &=& 
\epsilon_{\alpha \beta} s_\perp^\alpha P_{J\perp}^\beta
\frac{\alpha_s^2}{s} \sum_{a,b}  
\int \frac{dx}{x}\frac{dx'}{x'}f_{b/B}(x') 
\nnu
&&\times
\left[T_{a,F}(x, x) - x\frac{d}{dx}T_{a,F}(x, x)\right]
\nnu
&&
\times
\frac{1}{\hat{u}} H^{\rm Sivers}_{ab\to c}(\hat s,\hat t,\hat u)\delta\left(\hat s+\hat t+\hat u\right),
\label{pcs_twist3}
\een
where $\sum_{a,b}$ runs over all parton flavors, $f_{b/B}(x')$ is the collinear PDF in the unpolarized proton, and $\hat s$, $\hat t$, and $\hat u$ are the standard partonic Mandelstam variables~\cite{Kouvaris:2006zy,Kang:2012xf}. 
$H^{\rm Sivers}_{ab\to c}$ represent the cut scattering amplitudes 
 for the partonic process $a b\to c d$ with the expressions given in~\cite{Gamberg:2010tj,Kouvaris:2006zy}.   It is important to emphasize  that in 
the twist-3 collinear factorization approach, the process-dependence 
of the ISIs  and FSIs, which are determined from the color factors 
coming from the partonic process cut scattering amplitudes
 are absorbed into the 
short-distance perturbative hard-part functions, 
while the relevant twist-3 three-parton correlation 
functions $T_{q, F}(x, x)$ are universal or process independent.
It is because of this fact that the universal $T_{q, F}(x, x)$ is uniquely related to the Sivers function in SIDIS as in Eq.~\eqref{TF}. 
Thus, the process-dependence of the Sivers effect for jet production is included in $H^{\rm Sivers}_{ab\to c}$. Now since the 
SIDIS Sivers asymmetry is only associated with FSIs, while the jet spin asymmetry is associated with both ISIs and FSIs, by comparing the SIDIS measurement 
and the jet spin asymmetry, we are essentially testing the central role of 
these ISIs and FSIs, hence the {\it process dependence} of the Sivers effect.  
The jet SSA, $A_N$ is computed from the ratio of the spin-dependent to the spin-averaged cross section,
\ben
A_N = \left.E_J\frac{d\Delta\sigma(s_\perp)}{d^3P_J} \right/ E_J\frac{d\sigma}{d^3P_J},
\label{ssaAn}
\een
where the spin-averaged differential cross section $E_J\frac{d\sigma}{d^3P_J}$ in the denominator is defined in Ref.~\cite{Kang:2012xf}. 
\bef
\psfig{file=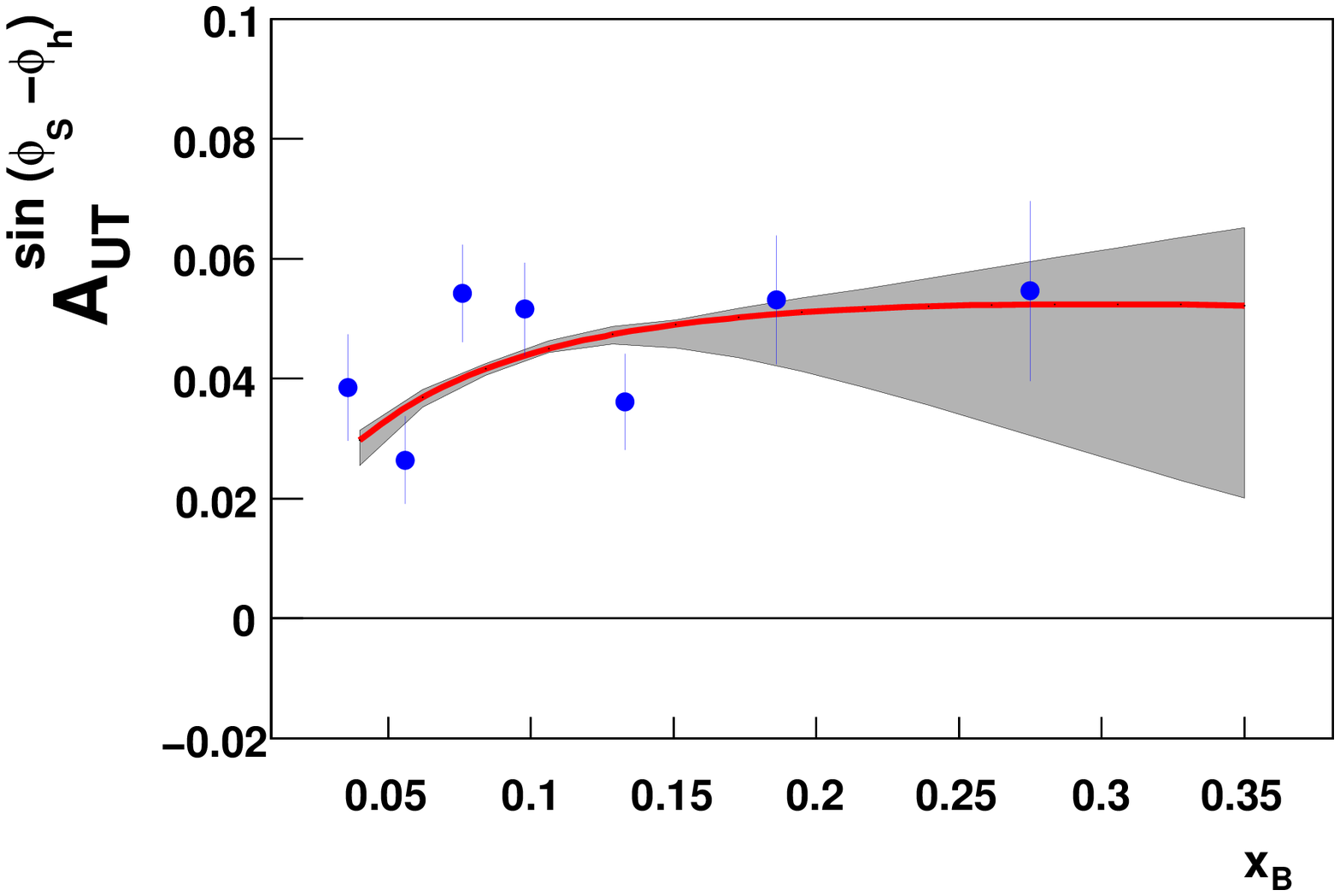, width=0.4\textwidth}
\caption{Description of the HERMES~\cite{hermes} data for $\pi^+$ production as a function of Bjorken $x_B$. The solid lines are the central values from Table~\ref{fitpar} and the shaded region corresponds to the parameter scan as explained in the text.}
\label{hermesfit}
\eef

To see whether the inclusive jet data in $pp$ collisions are consistent with the Sivers asymmetry data in SIDIS processes, we perform a global fit of the SIDIS Sivers asymmetry data collected by the HERMES and COMPASS experiments~\cite{hermes,compass} to extract the Sivers functions.   We then derive the functional form for twist-3 ETQS function $T_{q,F}(x,x)$ with the help of Eq.~\eqref{TF} and in turn compute the jet spin asymmetry $A_N$ from Eq.~\eqref{ssaAn} to be compared with the data collected by AnDY experiment~\cite{AnDY}. 

We adopt the Gaussian forms in Ref.~\cite{Anselmino:2008sga} for the spin-averaged PDFs, $f_{a/A}(x, k_\perp^2)$ and FFs $D_{h/a}(z, p_T^2)$, with the Gaussian width, $\langle k_\perp^2\rangle=0.25$ GeV$^2$ and $\langle p_T^2\rangle=0.2$ GeV$^2$. The quark Sivers function 
$f_{1T}^{\perp q}(x,k_\perp^2)$ for SIDIS is parameterized as, 
\ben
f_{1T}^{\perp q}(x, k_\perp^2) =- \, {\cal N}_q(x)  h(k_\perp) f_{q/A} (x, k_\perp^2),
\een 
where the  $k_\perp$-dependence $h(k_\perp) = \sqrt{2e}\,\frac{M}{M_{1}}\,e^{-{k}_\perp^2/{M_{1}^2}}$, with $M$ the proton mass, and the $x$-dependent coefficient ${\cal N}_q(x) =  N_q x^{\alpha_q}(1-x)^{\beta_q}(\alpha_q+\beta_q)^{\alpha_q+\beta_q}/({\alpha_q}^{\alpha_q}{\beta_q}^{\beta_q})$. 
For the purpose of our fit, we use GRV98LO for the spin-averaged collinear PDFs~\cite{Gluck:1998xa} and DSS parametrization for collinear FFs~\cite{deFlorian:2007aj}.  During the fit we enforce positivity bounds~\cite{Bacchetta:1999kz} on Sivers functions of quarks and anti-quarks. Thus, we 
will present our results separately as parametrizations for ``valence'' Sivers functions ($u_v$ and $d_v$) and ``sea'' Sivers functions ($\bar u$ and $\bar d$) in the end. We emphasize that in order to explore the uncertainty in Sivers function in the high-$x$ region, we allow $\beta_{u_v}$ and $\beta_{d_v}$ to vary independently as compared with the fit in Ref.~\cite{Anselmino:2008sga}.
\begin{table}[htb]
\centering
\caption{Best values of the free parameters  for the Sivers function from fit to SIDIS data~\cite{hermes, compass} on $A_{UT}^{\sin(\phi_h-\phi_s)}$.}
\label{fitpar}
\begin{tabular}{l l l l l l}
\hline
\hline
&&$\chi^2/d.o.f. = 1.04$&&&\\
\hline
$\alpha_{u_v}$ &=& $0.05^{+0.2}_{-0.05}$ & $\alpha_{d_v}$ &=& $0.76^{+0.20}_{-0.20} $  \\
$\beta_{u_v}$&=& $0.78^{+2.35}_{-0.77}$ & $\beta_{d_v}$ &=& $2.09^{+1.20}_{-0.90} $ \\
$N_{u_v}$ &=& $0.34^{+0.04}_{-0.04}$ & $N_{d_v}$ &=& $-1^{+0.42}_{-0} $ \\
$\alpha_{sea}$ &=& 0 fixed & $\beta_{sea}$ &=& 0  fixed \\
$N_{\bar u}$ &=&  $0.003^{+0.05}_{-0.05}$ & $N_{\bar d}$ &=&   $-0.15^{+0.08}_{-0.09}$\\
$M_1^2$ &=& $0.45^{+0.53}_{-0.22}$ (GeV/$c)^2$ & & & \\
\hline
\hline
\end{tabular}
\end{table}

Fitting the pion data from both HERMES and COMPASS we obtain a very good description of SIDIS data, with $\chi^2/d.o.f. = 1.04$. The resulting set of parameters are presented in Table~\ref{fitpar} together with the corresponding errors. As one can see, the biggest uncertainty is on parameters $\beta_{u_v}$ and $\beta_{d_v}$. This happens because SIDIS data covers a rather limited kinematic region in $x\lesssim 0.3$, as seen clearly in the HERMES plot Fig.~\ref{hermesfit}. Note that future measurements of JLab 12~\cite{Dudek:2012vr} will explore the high-$x$ region in SIDIS which is very important for jet $A_N$ as far as integration over $x$ is performed in Eq.~\eqref{pcs_twist3}.

In order to find the region of allowed values of $\beta_{u_v}$ and $\beta_{d_v}$, we  perform the scan procedure, also used in Ref.~\cite{Anselmino:2012rq} to study the Collins effect. We  produce a 
grid of values $\beta_{u_v}$, $\beta_{d_v}$ $\in [0,4]$ in  steps of 0.25 and for each pair of $\beta_{u_v}$, $\beta_{d_v}$ perform a fit of SIDIS data. The resulting sets of parameters corresponding to 289 pairs of $\beta_{u_v}$, $\beta_{d_v}$  give very good description of SIDIS data with $\chi^2/d.o.f \in [1.04,1.08]$;  they are all almost statistically identical. Using these 289 sets of parameters we draw the shaded corridor in all the plots.  It is important to 
realize that this corridor corresponds to almost the same  description of SIDIS data. 

\bef
\psfig{file=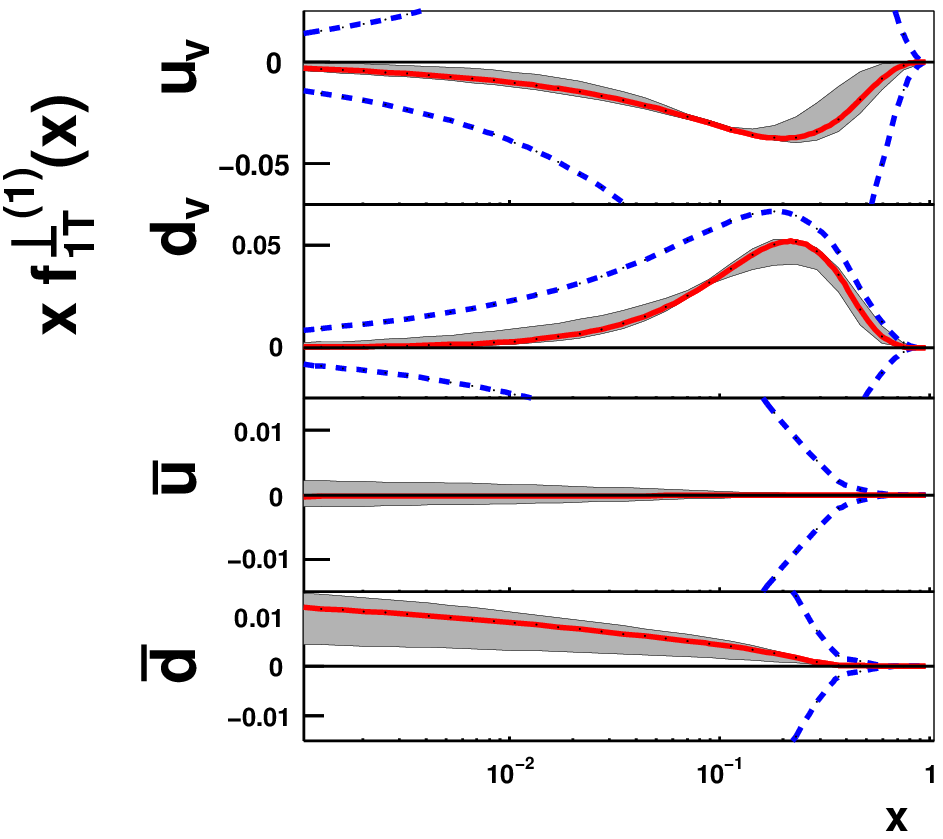, width=0.4\textwidth}
\caption{The first $k_{\perp}$-moment of the quark Sivers functions as a function of $x$, here $f_{1T}^{\perp q(1)}(x)=-T_{q,F}(x,x)/2M$. Dashed lines correspond to positivity bounds, while the solid lines and the shaded region are the same as in Fig.~\ref{hermesfit}.}
\label{siversfn}
\eef

We present a  comparison to the SIDIS data in Fig.~\ref{hermesfit}, which gives a very good description of HERMES $\pi^+$ data. For $\pi^0$, $\pi^-$ asymmetries and $z_h$ and $P_{h\perp}$ dependencies (and COMPASS data), the description is similar. In Fig.~\ref{siversfn} we show the first $k_\perp$-moment of the extracted quark Sivers functions versus $x$. 
\bef
\psfig{file=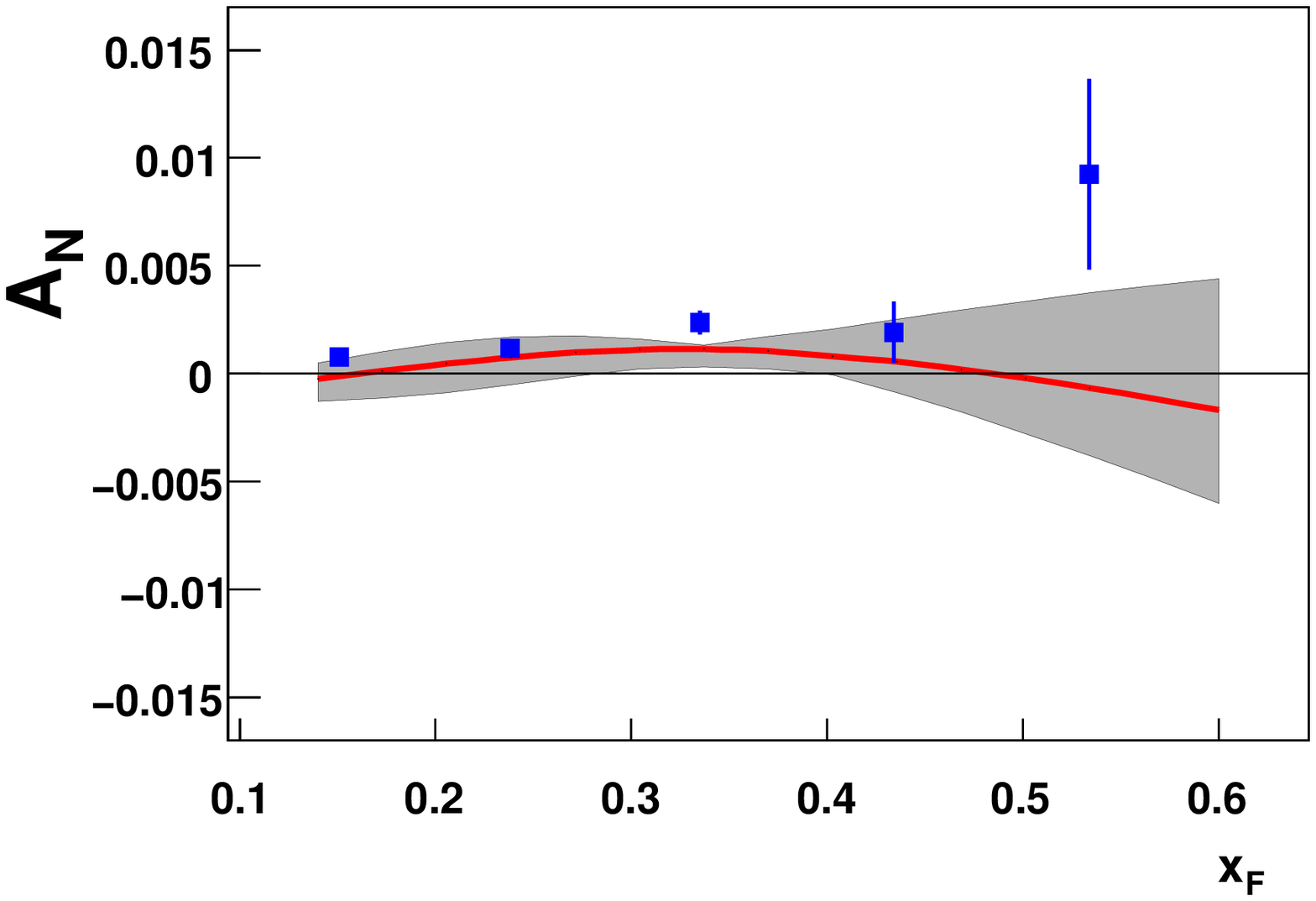, width=0.4\textwidth}
\caption{Description of AnDY data~\cite{AnDY} for inclusive jet production at forward rapidity $\langle y\rangle=3.25$ and  center-of-mass energy $\sqrt{s}$=500 GeV. Shaded region corresponds to the parameter scan.}
\label{jetan}
\eef

\bef
\psfig{file=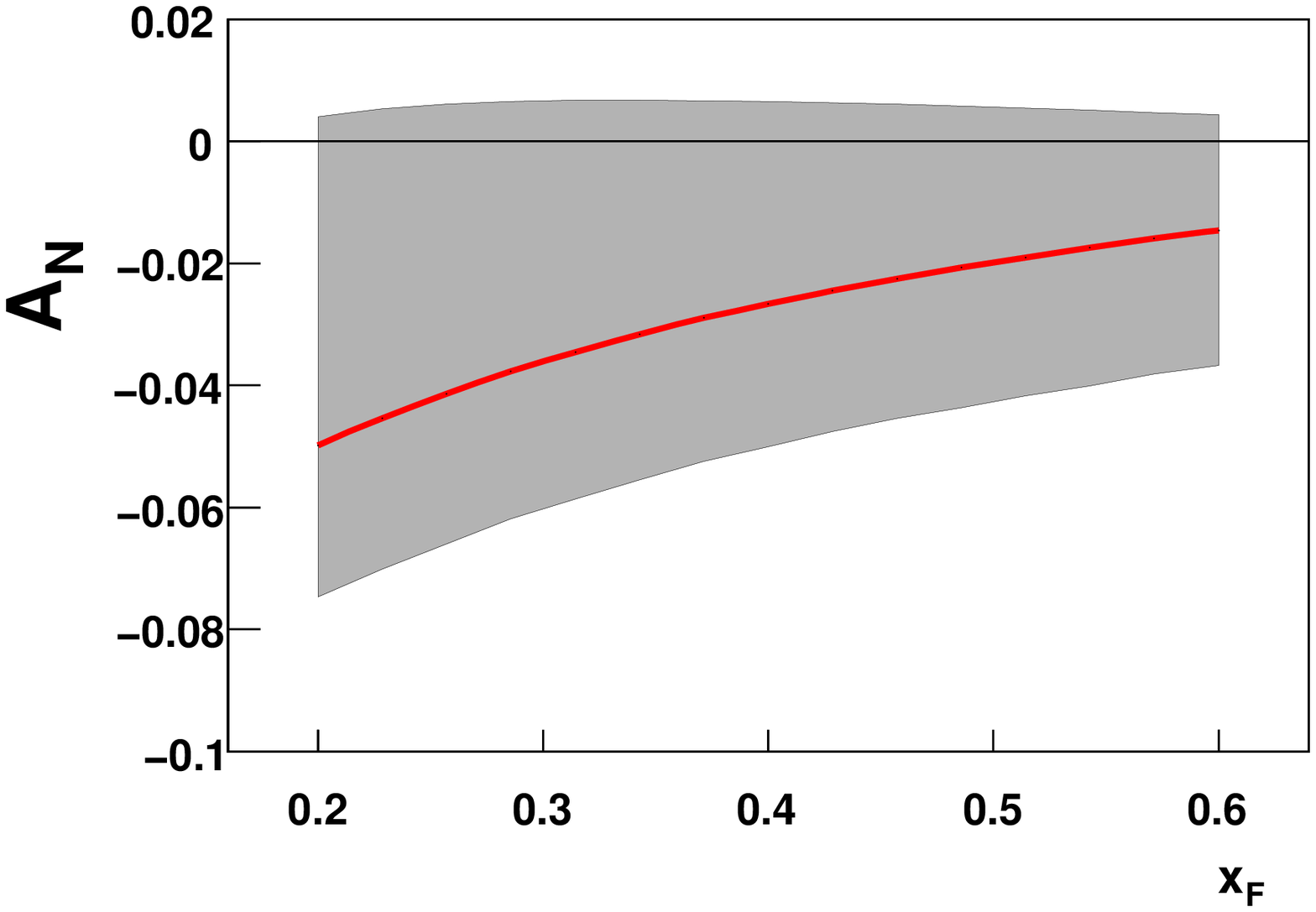, width=0.4\textwidth}
\caption{Prediction of direct photon $A_N$ in $pp$ collisions at rapidity $y=3.5$ and center-of-mass energy $\sqrt{s}$=200 GeV.}
\label{photonan}
\eef

We now assess whether the recently measured jet spin asymmetry from the AnDY experiment is compatible with the SIDIS Sivers asymmetry data; in other words, whether the jet asymmetry is consistent with our expectation on the process-dependence of the Sivers effect. To this end, we calculate the jet asymmetry $A_N$ from Eq.~\eqref{ssaAn} with our 289 equally-good sets of parameters. The resulting shaded region for jet $A_N$ as a function of Feynman $x_F\equiv 2P_{Jz}/\sqrt{s}$ ($P_{Jz}$ the jet longitudinal momentum) is shown in Fig.~\ref{jetan}. We note that the jet data are inside the 
shaded region, which demonstrates that SIDIS Sivers  data and jet $A_N$ data are statistically compatible with each other and that there exists a set of Sivers functions which can describe them simultaneously~\cite{Torino}. Although we 
cannot claim that our analysis proves the process-dependence of the Sivers effect (the role of ISIs and FSIs) 
due to the large uncertainty and the very small size of the jet asymmetry data, at the very least such a process-dependence is not in disagreement with the existing experimental data. 
Thus, we conclude that this is the first indication for the process-dependence of the Sivers effect. 

As far as the jet is concerned, since it is produced through strong interaction initiated processes, in contrast to  direct photon, SIDIS, or DY, 
the very small size of the jet asymmetry is largely due to a cancellation 
between $u$ and $d$ quark Sivers functions (which have opposite signs, an observation from the SIDIS fit and also from the prediction based on
large-$N_c$ QCD \cite{Pobylitsa:2003ty}) which are neither weighted by 
the quark electric charges squared nor coupled to fragmentation functions.
  In order to carry out a more definitive 
test on the process-dependence of the Sivers function, and 
simultaneously explore the consistency of the 
TMD and collinear twist-3 factorization formalisms, it is 
advantageous to study hadronic processes that have 
larger spin asymmetries. In this respect DY production is the  
ideal process to explore process dependence, provided that the effects of TMD evolution are completely understood; while direct photon production (though experimentally challenging~\cite{Bland}) can be used to study the consistency of the factorization formalisms. First we make 
a prediction for the spin asymmetry $A_N$ for direct photon production at RHIC kinematics, $y=3.5$ and $\sqrt{s}=200$ GeV, in Fig.~\ref{photonan}. Since $u$ and $d$ quark Sivers functions are now weighted with their electric charge squared, which compensates the cancellation between them, we found that the direct photon $A_N$ has much larger size $\sim 5\%$, and it is negative~\cite{pion-pp} due to the nature of ISIs associated with the Sivers effect for direct photon production. 

Now, using the TMD factorization formalism~\cite{Drell-Yan1, Drell-Yan10},  we compute the 
DY spin asymmetry $A_N$ at center-of-mass energy $\sqrt{s}=500$ GeV, for invariant mass $4<Q<8$ GeV and transverse momentum $0<q_\perp<1$ GeV. Due to the same reason, the asymmetry is larger $\sim 8\%$ and negative~\cite{Drell-Yan2} (see Fig.~\ref{DY}).  At small and intermediate $x_F$ region, the behavior is very similar to those in~\cite{Drell-Yan1, Drell-Yan10}. The large $x_F$ behavior is different though it has a similar uncertainty band. 
\bef
\psfig{file=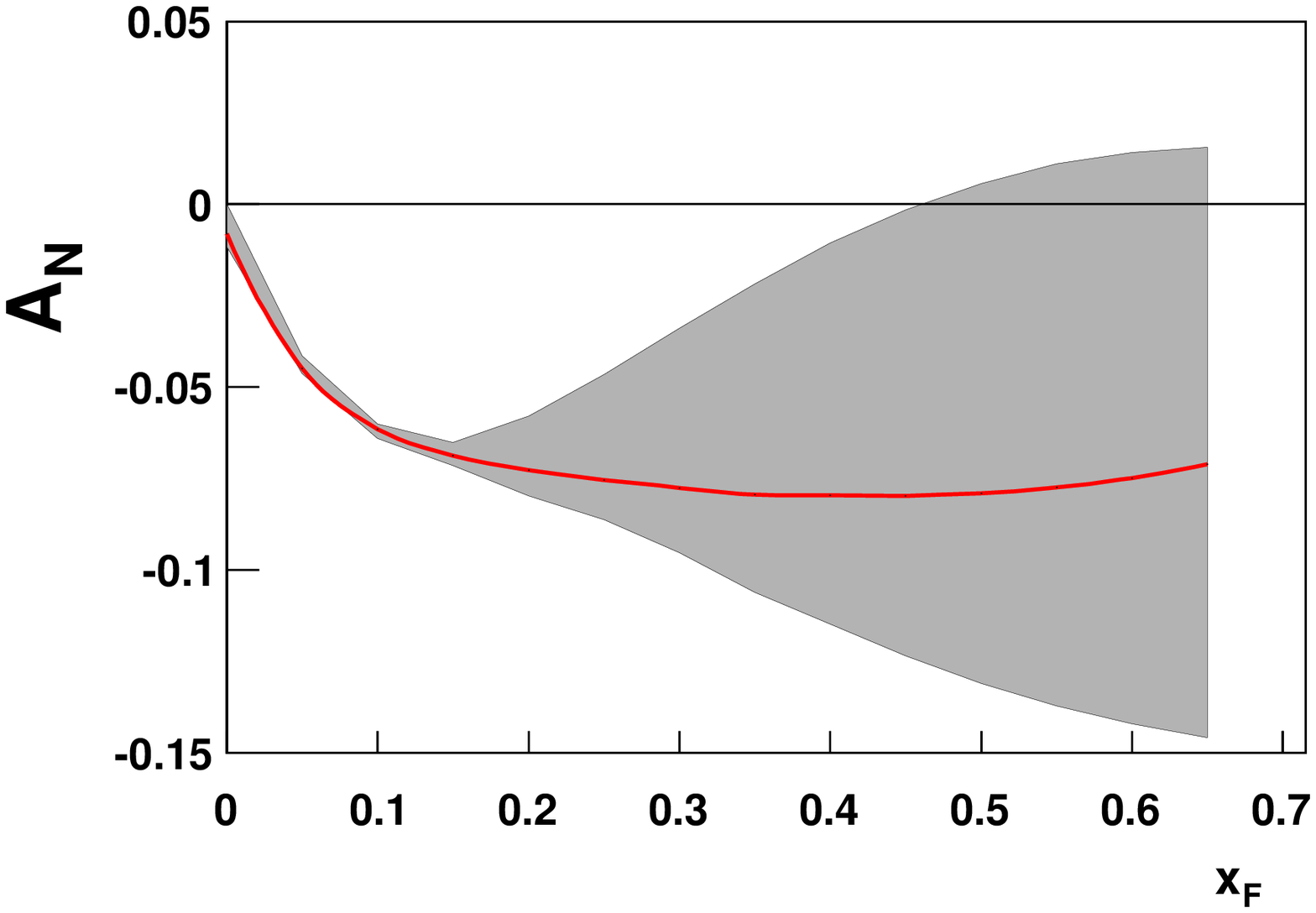, width=0.4\textwidth}
\caption{Prediction for Sivers asymmetry $A_N$ for DY lepton pair production at $\sqrt{s}=500$ GeV, for the invariant mass $4<Q<8$ GeV and transverse momentum $0<q_\perp<1$ GeV. Note our $A_N$ is related to the weighted asymmetry defined in Ref.~\cite{Drell-Yan1} as $A_N=-A_{UT}^{\sin(\phi_{\gamma}-\phi_S)}$.}
\label{DY}
\eef

In summary, we have analyzed the SSA for inclusive jet production in $pp$ collisions collected by AnDY experiment and the Sivers asymmetry data from SIDIS experiments. We study the effect of ISIs and FSIs between the active parton and the remnant of the hadron on the process-dependence of the Sivers effect for both processes. After carefully taking into account 
all ISIs and FSIs in the inclusive jet production, we find that 
the calculated jet spin asymmetry based on the Sivers function extracted from SIDIS experiments are consistent with the AnDY experimental data. 
Our result provides a first indication for the process-dependence of the Sivers effect and further demonstrates consistency between the TMD and collinear twist-3 factorization formalisms. However, due to the large uncertainty of the 
current data from AnDY and the small size of the jet spin asymmetry, our result cannot provide  conclusive confirmation for  
process-dependence. Thus we also propose direct photon spin asymmetry
along with DY measurements 
to test the process dependence of the Sivers effect. They are complementary to each other, 
DY production is the  
ideal process to explore process dependence, provided that the effects of TMD evolution are completely understood; while   
direct photon production can be used to study the consistency of the factorization formalisms.
We make predictions for these asymmetries at RHIC kinematics, 
and encourage RHIC to undertake both measurements.

\begin{acknowledgments}
We thank M. Anselmino, L. Bland,  H. Crawford, K.O. Eyser, R. Fatemi, C. Perkins, W. Vogelsang, and F. Yuan for helpful discussions.  This work is supported by the U.S. Department of Energy under Contract Nos. DE-FG02-07ER41460 (L.G.), DE-AC02-05CH11231 (Z.K.), and DE-AC05-06OR23177 (A.P.).
\end{acknowledgments}


\begin{thebibliography}{99}
\bibitem{Drell-Yan2}
Note here we neglect the effect of QCD evolution, which we plan to investigate in a future publication.
\end{thebibliography}
\end{document}